\documentclass[prl,twocolumn,amsmath,amssymb,superscriptaddress]{revtex4}

\usepackage{graphicx}
\usepackage{dcolumn}
\usepackage{bm}
\usepackage{ textcomp }
\usepackage{color}
\usepackage{amsmath}
\usepackage{amssymb}
\usepackage{xcolor}
\usepackage{hyperref}
\usepackage{easyReview}
\usepackage{mathdots}
\usepackage{float}
\usepackage{lineno}
\usepackage{array}
\usepackage{xcolor,colortbl}
\definecolor{LightBlue}{rgb}{0.8,0.8,0.8}

\usepackage[normalem]{ulem}

\UseRawInputEncoding

\begin{document}
\title{Phase transitions of the ferroelectric $(\mathrm{ND}_4)_2\mathrm{FeCl}_5\cdot\mathrm{D}_2\mathrm{O}$ under a magnetic field}
\date{\today}

\author{Shaozhi Li}
\affiliation{Materials Science and Technology Division, Oak Ridge National Laboratory, Oak Ridge, Tennessee 37831, USA}
\author{Randy S. Fishman}
\affiliation{Materials Science and Technology Division, Oak Ridge National Laboratory, Oak Ridge, Tennessee 37831, USA}

\begin{abstract}
Due to the strong coupling between magnetism and ferroelectricity,
$(\mathrm{ND}_4)_2\mathrm{FeCl}_5\cdot\mathrm{D}_2\mathrm{O}$ exhibits several intriguing magnetic and electric phases. 
In this letter, we include high-order onsite spin anisotropic interactions in a spin model that successfully captures the ferroelectric phase transitions of $(\mathrm{ND}_4)_2\mathrm{FeCl}_5\cdot\mathrm{D}_2\mathrm{O}$ under a magnetic field and produces the large weights of high-order harmonic components in the cycloid structure that are observed from neutron diffraction experiments. Moreover, we predict a new ferroelectric phase sandwiched between the FE II and FE III phases in a magnetic field.
Our results emphasize the importance of the high-order spin anisotropic interactions and provide a guideline to understand multiferroic materials with rich phase diagrams.
\end{abstract}

\maketitle
{\it Introduction.} --- The interplay between charge, lattice, and spin degrees of freedoms induces many fascinating phenomena in materials, including  multiferroic behavior~\cite{Hill,Tokura_2014,spaldin2019,WangNPG}, colossal magnetoresistance~\cite{Qiu2018,Baldini10869} and stripe order in the cuprates~\cite{Zhang68,ZhaoNatureMaterials2019} and nickelates~\cite{ZhangPRL2019,Merritt2020}.
In the past few years, significant progress has been made in understanding and discovering multiferroics~\cite{Daniel,Dong2015,Fiebig2016}, motivated by the promise of new technological applications in energy transformation and signal generation and processing. In general, there are two types of multiferroic materials: type I, where the ferroelectricity does not originate from the magnetic order~\cite{Wang1719}, and type II, which is more interesting because the electric polarization appears as a consequence of the magnetic order\cite{Kimura2003,Daniel,XiangPRL2008,MalashevichPRL,KenzelmannPRL,SolovyevPRB,SergienkoPRB2006}.

Recently, a new type II material $(\mathrm{NH}_4)_2\mathrm{FeCl}_5\cdot\mathrm{H}_2\mathrm{O}$ with a rich 
phase diagram was discovered~\cite{TianPRB2016,TianPRB2018,Ackermann_2013}. $(\mathrm{NH}_4)_2\mathrm{FeCl}_5\cdot\mathrm{H}_2\mathrm{O}$ has an incommensurate cycloidal magnetic order in the $ac$ plane with wave vector ${\bf Q}=(0,0,0.23)$ r.l.u. below 6.9~K~\cite{Alberto2015,TianPRB2016}. An incommensurate sinusoidal collinear state appears between 6.9~K and 7.5~K~\cite{TianPRB2016}. Ferroelectricity below 6.9~K is attributed to the inverse Dzyaloshinskii-Moriya (DM) mechanism, which predicts that the electric polarization is proportional to $({\bf S}_i\times{\bf S}_j)\times{\bf Q}$, leading to an electric polarization along the $a$-axis~\cite{Alberto2015}. These properties have been extensively discussed in previous inelastic neutron scattering experiments~\cite{Xiaojianpre} and theoretical studies based on density functional theory (DFT)~\cite{Clunenpj} and spin models~\cite{Minseong2021}.

\begin{figure}[t]
\center\includegraphics[width=0.99\columnwidth]{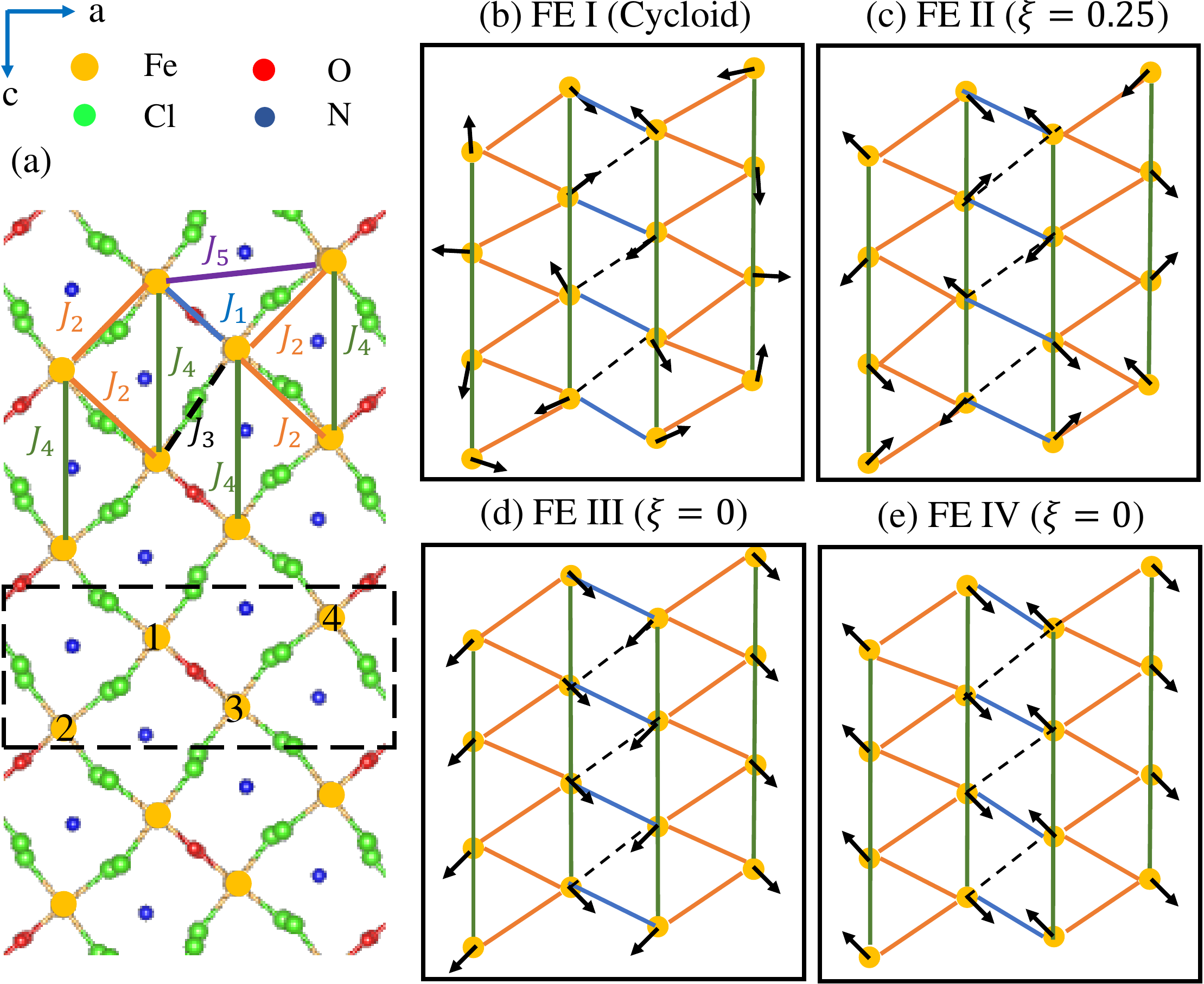}
\caption{\label{Fig:fig1} {Crystal structure and magnetic order.} (a) Crystal structure of $(\mathrm{ND}_4)_2\mathrm{FeCl}_5\cdot\mathrm{D}_2\mathrm{O}$ in the $ac$ plane. Five different exchange interactions are labeled with  different colors. The dashed-rectangle represents one unit cell with four Fe atoms labeled by Arabic numbers. (b) - (e) Spin configurations for four different ferroelectric phases.
}
\end{figure}

One exciting feature of $(\mathrm{NH}_4)_2\mathrm{FeCl}_5\cdot\mathrm{H}_2\mathrm{O}$ is that the direction of the electric polarization can be tuned by a magnetic field~\cite{Ackermann_2013}. When the magnetic field is applied along the $a$-axis at low temperature, phase transitions from ferroelectric I (FE I) to ferroelectric II (FE II) to ferroelectric III (FE III) phases are observed. 
Neutron diffraction measurements show that the magnetic wave vector smoothly increases with the magnetic field in FE I, jumps to ${\bf Q}=(0,0,0.25)$ r.l.u. in FE II, and then to ${\bf Q}=(0,0,0)$ r.l.u. in FE III~\cite{AlbertoPRB2017} (see Fig.~\ref{Fig:fig1}). 
The critical magnetic fields for these two transitions are about 2.8~T and 4.7~T near zero temperature, respectively. 
In FE I and FE II, the electric polarization lies along the $a$-axis;  in FE III, the electric polarization rotates to the $c$-axis.
It has been proposed that the microscopic mechanism of multiferroicity changes from the 
inverse DM interaction in FE I and FE II to $p$-$d$ hybridization in FE III.
While these phase transitions are also observed when the magnetic field is applied along the $c$-axis, the critical fields become 1.3~T and 2.2~T. The different critical magnetic fields along $a$- and $c$-axes imply that the spin is not isotropic in the $ac$ plane.

Knowledge of the spin structure in $(\mathrm{NH}_4)_2\mathrm{FeCl}_5\cdot\mathrm{H}_2\mathrm{O}$ is limited to zero field. 
To understand the spin behavior of $(\mathrm{NH}_4)_2\mathrm{FeCl}_5\cdot\mathrm{H}_2\mathrm{O}$ under a magnetic field, we need to carefully consider the effect of spin anisotropy.  Previous theoretical investigations use a simplified form for the anisotropy that cannot explain the spin behavior under a magnetic field.
In this letter, we 
address this issue by studying a Heisenberg Hamiltonian with two onsite anisotropic interactions.
Examining the second-order onsite anisotropy reveals that this interaction alone cannot describe the spin behavior of $(\mathrm{ND}_4)_2\mathrm{FeCl}_5\cdot\mathrm{D}_2\mathrm{O}$, including the 
strong observed intensities of the third and fifth harmonic components of the cycloidal state and the appearance of the FE II state under a magnetic field.  Interestingly, these deficiencies can be addressed by adding a fourth-order onsite anisotropy. Our results imply that the high-order anisotropic interactions are crucial to explain the spin properties of $(\mathrm{ND}_4)_2\mathrm{FeCl}_5\cdot\mathrm{D}_2\mathrm{O}$.


\begin{figure}[t]
\center\includegraphics[width=\columnwidth]{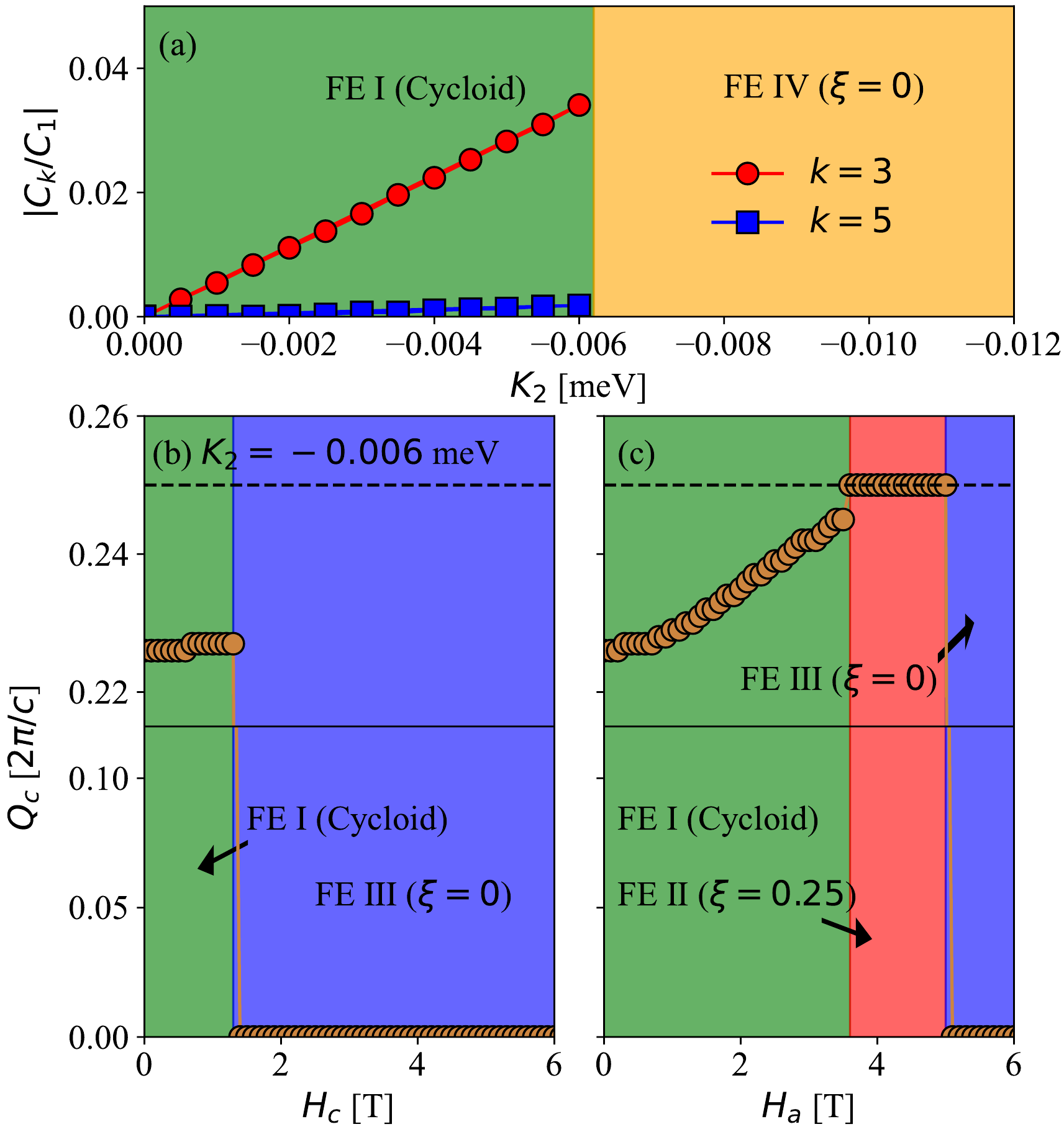}
\caption{\label{Fig:fig2} {Phase diagrams at $K_4=0$.} Panel(a) plots $|\frac{C_3}{C_1}|$ and $|\frac{C_5}{C_1}|$ as a function of $K_2$. Panels (b) and (c) plot the evolution of the wave vector $Q_c$ as a function of the magnetic fields $H_c$ and $H_a$, respectively. 
}
\end{figure}

\begin{figure*}[t]
\center\includegraphics[width=0.9\textwidth]{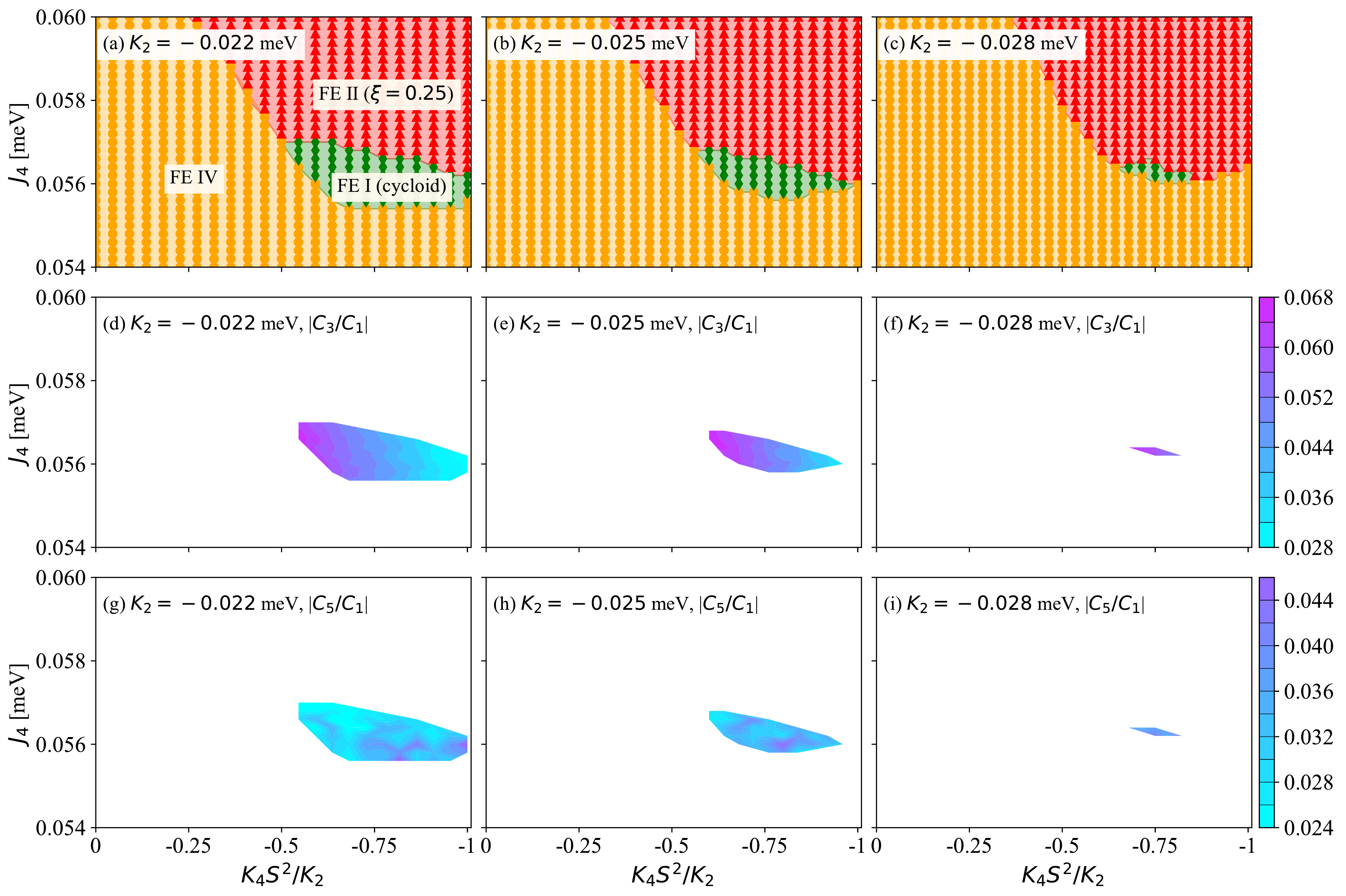}
\caption{\label{Fig:fig3} {Phase diagrams for $K_2=-0.22$, $-$0.025, and $-$0.028~meV.} (a) - (c) Phase diagrams in the plane of $J_4$ and $K_4S^2/K_2$ for three different $K_2$ values. (d) - (f) $|C_3/C_1|$ in the cycloidal state for three different $K_2$ values. (g) - (i) $|C_5/C_1|$ in the cycloidal state for three different $K_2$ values.  
}
\end{figure*}

\begin{figure}[t]
\center\includegraphics[width=0.99\columnwidth]{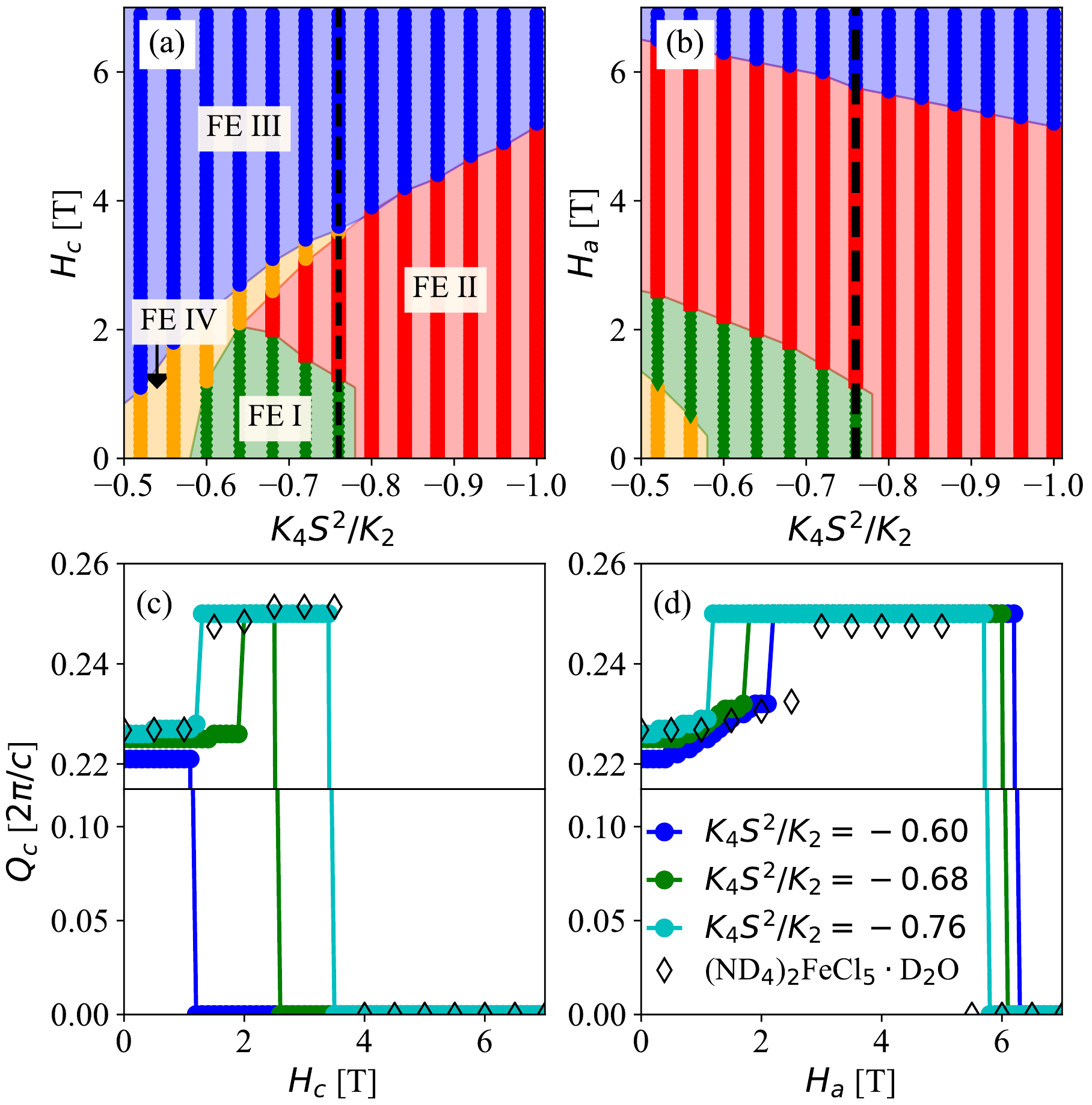}
\caption{\label{Fig:fig4} {Phase transitions under the magnetic field at ${J_4=0.0566}$~meV and ${K_2=-0.025}$~meV.} Panels (a) and (b) plot the phase diagram in the $H_c$ and $K_4$ plane and in the $H_a$ and $K_4$ plane, respectively. Panels (c) and (d) plot the change of the magnetic wave vector $Q_c$ as a function of $H_c$ and $H_a$, respectively. Diamond symbols represent experimental results obtained from Ref.~\cite{AlbertoPRB2017}.
}
\end{figure}

\begin{figure}[t]
\center\includegraphics[width=0.99\columnwidth]{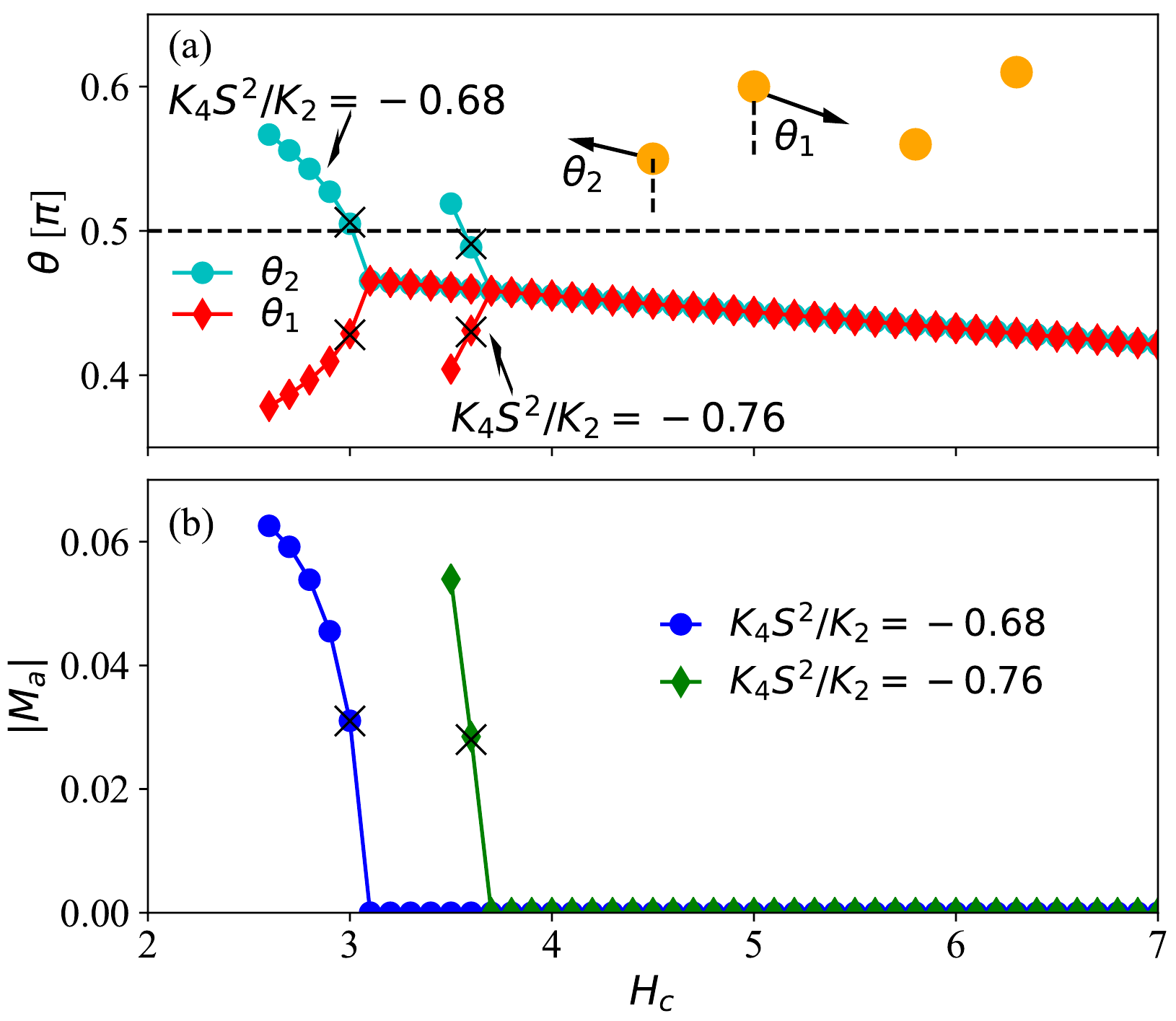}
\caption{\label{Fig:fig5} {Comparisons of FE III and FE IV.} Panel (a) plots spin angles of two neighboring sites along the $c$-axis. Panel (b) plots the $a$ component of the uniform magnetization $M_{a}$. Crossing symbols represent the phase boundary between FE III and FE IV. Here, $J_2$ and $K_2$ have the same values as those in Fig.~\ref{Fig:fig4}. 
}
\end{figure}

{\it Magnetic Anisotropy.} --- In general, the magnetic anisotropy in solids is induced by the spin-orbit coupling~\cite{JiaJCP2014,DaiJCC2008,ShaozhiPRB2021}, which is given by
$H^\prime=\lambda \mathbf{S}\cdot \mathbf{L} - \mu_B \mathbf{S}\cdot \mathbf{H} - 2\mu_B \mathbf{L}\cdot \mathbf{H}$.
Here, $\lambda$ is the spin-orbit coupling strength, $\mathbf{S}$ and $\mathbf{L}$ represent the spin and angular momentum operators, and $\mathbf{H}$ is the magnetic field. Integrating over the angular momentum operator in the atomic limit, the second-order perturbative energy is $E^{(2)}\propto -\lambda^2 \Lambda_{\alpha\beta} S_{\alpha}S_{\beta} + 2\mu_B(\delta_{\alpha\beta}-\Lambda_{\alpha\beta})S_{\alpha}H_{\beta}$, where $\alpha$ and $\beta$ are indexes for $x$, $y$, and $z$.
By considering crystal symmetry, $\lambda^2 \Lambda_{\alpha\beta} S_{\alpha}S_{\beta}$ can be reduced to $K_2 S_z^2$ in $\mathrm{SrFeO}_2$~\cite{XiangPRL2008}, $\mathrm{Sr}_3\mathrm{Fe}_2\mathrm{O}_5$~\cite{Koo2009}, $\mathrm{TbMnO}_3$~\cite{XiangPRL2008}, and $\mathrm{Ag}_2\mathrm{MnO}_2$~\cite{JiSPRB2010}. The term $2\mu_B(\delta_{\alpha\beta}-\Lambda_{\alpha\beta}S_{\alpha}H_{\beta})$ induces anisotropic $g$-factors,  which can account for the spin anisotropy in $\mathrm{Tb}_2\mathrm{Ir}_3\mathrm{Ga}_9$~\cite{FengYearxiv2021}. 
If the spin-orbit coupling $\lambda$ is strong, the fourth-order perturbative energy $E^{(4)}\propto -\lambda^4 U_{\alpha\beta\gamma\eta}S_{\alpha}S_{\beta}S_{\gamma}S_{\eta}$ must also be considered.

In this letter, we study the effect of the onsite anisotropic interaction $K_2 S_a^2 +K_4S_a^4$ in $(\mathrm{ND}_4)_2\mathrm{FeCl}_5\cdot\mathrm{D}_2\mathrm{O}$, where $K_2 S_a^2$ and $K_4 S_a^4$ originate from the second-order and fourth-order perturbative terms. The full spin Hamiltonian for $(\mathrm{ND}_4)_2\mathrm{FeCl}_5\cdot\mathrm{D}_2\mathrm{O}$ is given by
\begin{eqnarray}
H&=&\sum_{i,j}J_{i,j}\boldsymbol{S}_i\cdot \boldsymbol{S}_j + D\sum_i(S_{i,b})^2+K_2\sum_i(S_{i,a})^2 \nonumber\\ && + K_4\sum_i(S_{i,a})^4+g\mu_B\sum_i(H_a S_{i,a} + H_c S_{i,c}),
\end{eqnarray}
where $\boldsymbol{S}_i$ is the spin operator of the $\mathrm{Fe}^{3+}$ ion on site $i$ with length $S=5/2$ and $J_{i,j}$ is the exchange interaction, which is labeled in Fig.~\ref{Fig:fig1}(a). $D$, $K_2$, and $K_4$ are the single-ion anisotropic interactions.  If not stated otherwise, we use previous INS studies~\cite{Xiaojianpre} to set $\{J_1,J_2,J_3,J_4,J_5\}=\{0.178,0.0641,0.0289,0.0566,0.0447\}$~meV and $D=0.0183$~meV.

We use the variational method to study the Heisenberg model~\cite{FishmanPRB2013,supplement}. Three trial wave functions are used to obtain four different magnetically ordered states, which are labeled as FE I, FE II, FE III, and FE IV (see Fig.~\ref{Fig:fig1}). The magnetic wave vector is labeled as ${\bf Q}=(0,0,\xi)$ r.l.u.. FE I refers to the cycloidal state with $\xi<0.25$; FE II and FE IV are antiferromagnetic states with $\xi=0.25$ and $\xi=0$, respectively; FE III has spins that are canted by the magnetic field with $\xi=0$.
While both FE III and FE IV have $\xi=0$, FE III and FE IV are distinct states that appear at high and intermediate to low fields, respectively. 

{\it The second-order anisotropy.}--- We begin with the second-order interaction $K_2$ and set the fourth-order interaction $K_4$ to zero. Figure.~\ref{Fig:fig2}(a) shows the phase transition from FE I to FE IV as $K_2$ increases. The critical value for the phase transition is about $K_2=-0.0061$~meV. A polarized neutron diffraction experiment showed that the reflection intensities at $(0,0,3\xi)$ and $(0,0,5\xi)$ are about $I_3/I_1=0.0076$ and $I_5/I_1=0.0038$, implying that the cycloidal structure of $(\mathrm{ND}_4)_2\mathrm{FeCl}_5\cdot\mathrm{D}_2\mathrm{O}$ is strongly distorted with harmonics $|C_k| = C_1(I_k/I_1)^{1/2}$ associated with the $k{\bf Q}=(0,0,k\xi )$ r.l.u. components of the cycloid~\cite{TianPRB2016} (see supplementary material~\cite{supplement}). 
The maximum predicted values of $|C_3/C_1|$ ($I_3/I_1$) and $|C_5/C_1|$ ($I_5/I_1$) in FE I are about 0.038 (0.0014) and 0.0015 ($2.25\times 10^{-6}$), much smaller than the experimental results.

We now study the evolution of the spin structures under a magnetic field with $K_2=-0.006$~meV, which produces a relative large $|C_3/C_1|$ in the cycloidal state. Figures~\ref{Fig:fig2} (b) and~\ref{Fig:fig2} (c) plot the change of the wave vector $Q_c=\xi \frac{2\pi}{c}$ with the magnetic fields $H_c$ and $H_a$, respectively, where $c$ is the lattice constant along the $c$-axis.
When a field along $c$ ($H_c$) is applied, FE I directly transforms into FE III. When a field along $a$ ($H_a$) is applied, FE II appears between FE I and FE III.
In the cycloidal state, the wave vector $Q_c$ weakly depends on $H_c$, but it smoothly increases with $H_a$.

These theoretical results are inconsistent with the experiments in two respects. First, the third and fifth harmonic components are weak compared to experiments. Second, FE II is missing when the magnetic field is applied along the $c$-axis. 
These inconsistencies can be addressed by considering the fourth-order anisotropy.

{\it The fourth-order anisotropy.} --- To understand the fourth-order anisotropy, we calculate the phase diagrams in the plane of $J_4$ and $K_4S^2$ for three different values of $K_2$. Results are plotted in Fig.~\ref{Fig:fig3}, where solid symbols represent simulation results. The phase boundary lies at the middle of two data points. Figures~\ref{Fig:fig3}(a)-~\ref{Fig:fig3}(c) show that FE II is located on the right upper side and FE IV is located on the left bottom side. FE I resides between FE II and FE IV. As $|K_2|$ increases, the FE IV region
grows toward the right upper side, and the FE II region grows toward the left bottom side, thereby shrinking the FE I region. FE I disappears when $K_2<-0.028$~meV. 
Figures~\ref{Fig:fig3}(d)-~\ref{Fig:fig3}(f) and Figures~\ref{Fig:fig3}(g)-~\ref{Fig:fig3}(i) show the weight of the third ($|C_3/C_1|$) and fifth harmonic ($|C_3/C_1|$) components in the cycloidal state, respectively. $|C_3/C_1|$ smoothly decreases as $|K_4|$ increases, while there is no monotonic behavior for $|C_5/C_1|$. 
$|C_3/C_1|$ has a larger value ($\sim 0.06$) at the left upper corner in Figs.~\ref{Fig:fig3}(d)-~\ref{Fig:fig3}(i) compared to the other region, and $|C_5/C_1|$ is about 0.03 in that region.
While these results are still smaller than the experimental values, they are much larger than the results at $K_4=0$, especially with $|C_5/C_1|$ enhanced by a factor of ten.  It has been proposed that the strong spin-lattice interaction in $(\mathrm{ND}_4)_2\mathrm{FeCl}_5\cdot\mathrm{D}_2\mathrm{O}$~\cite{TianPRB2016} induces nonuniform spin-spin interactions that could further enhance the weight of the third and fifth harmonics.  It would be interesting to study the effect of the spin-lattice interaction in $(\mathrm{ND}_4)_2\mathrm{FeCl}_5\cdot\mathrm{D}_2\mathrm{O}$, but such a study is beyond our current focus.

Next, we study the phase transitions under a magnetic field. To be consistent with the microscopic model proposed in Ref.~\cite{Xiaojianpre}, we set $J_4=0.0566$\,meV. The value $K_2=-0.075$\,meV is used because it can produce relative large $|C_3/C_1|$ and $|C_5/C_1|$ in a wide region of $K_4$ at $J_4=0.0566$\,meV, as shown in Fig.~\ref{Fig:fig3}(e).  While a slightly different value of $K_2$ can quantitatively change results, the qualitative results remain the same. Figures~\ref{Fig:fig4}(a) and~\ref{Fig:fig4}(b) show the phase diagram in the $H_{c}$ and $K_4S^2/K_2$ plane and the $H_{a}$ and $K_4S^2/K_2$ plane, respectively. As was the case in Fig.~\ref{Fig:fig3}, solid symbols represent simulation results, and the phase boundary lies at the middle of two data points. Compared to Figs.~\ref{Fig:fig2}(b) and~\ref{Fig:fig2}(c), the phase diagram is richer when the fourth-order anisotropy is included.

At zero magnetic field, the magnetic ground state is FE IV for $K_4S^2/K_2>-0.58$, FE I for $-0.58>K_4S^2/K_2>-0.78$, and FE II for $-0.78>K_4S^2/K_2>-1$. When a field along $c$ is applied, both FE IV and FE II directly transform into FE III at high fields. For the FE I state, there are two different sets of phase transitions. When $-0.58>K_4S^2/K_2>-0.65$, FE I transforms to FE IV and then to FE III as $H_c$ increases; while for $-0.58>K_4S^2/K_2>-0.65$, FE I first transforms to FE II, then to FE IV, and finally to FE III.

The change of the magnetic wave vector $Q_c$ is different for these two sets of phase transitions. Figure~\ref{Fig:fig4}(c) plots the evolution of $Q_c$ under $H_c$ for three different values of $K_4S^2/K_2$. For $K_4S^2/K_2=-0.6$ (the former set of phase transitions), $Q_c$ is independent of $H_c$ in FE I, and jumps to zero in FE IV. For $K_4S^2/K_2=-0.68$ and $K_4S^2/K_2=-0.76$ (the latter set of phase transitions), $Q_c$ weakly depends on $H_c$ in FE I and then jumps to 0.25 $\frac{2\pi}{c}$ in
FE II. Finally, $Q_c$ is zero in FE IV and FE III. The change of $Q_c$ in the latter phase transitions is consistent with the results of the neutron diffraction measurements on $(\mathrm{ND}_4)_2\mathrm{FeCl}_5\cdot\mathrm{D}_2\mathrm{O}$~\cite{AlbertoPRB2017}, which are labeled as the diamond symbols in Figs.~\ref{Fig:fig4}(c) and~\ref{Fig:fig4}(d) .

Interestingly, the phase transition under $H_a$ is very different from that under $H_c$. Rather than transform directly into FE III, FE IV transforms continuously from FE I to FE II and then to FE III as $H_a$ increases. Compared to the case with field along $c$, FE IV is absent when $-0.58>K_4S^2/K_2>-0.78$. We also find that the critical value of $H_a$ for FE III is much larger than that of $H_c$; while the critical value of $H_a$ for the phase transition from FE I to FE II is close to the value of $H_c$ when $-0.64>K_4S^2/K_2>-0.78$. We plot the change of $Q_c$ with field $H_a$ in Fig.~\ref{Fig:fig4}(d). Here, the increase of $Q_c$ in FE I is more prominent than that in Fig.~\ref{Fig:fig4}(c) for the field along $c$.

In $(\mathrm{ND}_4)_2\mathrm{FeCl}_5\cdot\mathrm{D}_2\mathrm{O}$, 
FE II and FE III pop up near $H_c=1.5$ T and $H_c=4$ T or $H_a=2.7$ T and $H_a=5$ T, respectively~\cite{AlbertoPRB2017}. In our simulations, the critical values of $H_c$ ($H_a$) for these two states are 1.3 T (1.2 T) and 3.5 T (5.8 T) at $K_4S^2/K_2=-0.076$ and $K_2=-0.025$\,meV [see the dashed line in Figs.~\ref{Fig:fig4}(a) and ~\ref{Fig:fig4}(b)]. 
The small discrepancy between our theoretical and experimental results could originate from the change of exchange and anisotropy interactions in $(\mathrm{ND}_4)_2\mathrm{FeCl}_5\cdot\mathrm{D}_2\mathrm{O}$ in a magnetic field~\cite{Xiaojianpre}. 

Since both FE IV and FE III have zero wave vector, they cannot be distinguished based on $Q_c$.
Here, we compare FE III and FE IV and propose a method to distinguish them experimentally. We label the spin angles of neighbouring sites along the $c$-axis as $\theta_1$ and $\theta_2$, respectively, as shown in Fig.~\ref{Fig:fig5}(a).
While $\theta_1$ and $\theta_2$ are the same for FE III, they are different for FE IV.
Figure~\ref{Fig:fig5}(a) plots these two angles versus the field $H_c$ at $K_4S^2/K_2=-0.68$ and $K_4S^2/K_2=-0.76$. Here, $J_4=0.0566$~meV and $K_2=-0.025$~meV. Notice that $\theta_1$ increases and $\theta_2$ decreases as $H_c$ increases in FE IV and they are equal in FE III. We set the phase boundary between FE III and FE IV at a data point where $\theta_2$ crosses the dashed line $\theta=0.5\pi$ and label it as a crossing symbol in Fig.~\ref{Fig:fig5}. Figure~\ref{Fig:fig5}(b) shows the $a$ component of the magnetization $|M_a|$.
Under $H_c$, the spins of both FE III and FE IV are canted along the $c$ direction. However, $M_a$ is zero for FE III and it has a finite value for FE IV. As FE IV transforms into FE III, $|M_a|$ rapidly vanishes. Hence, FE III and FE IV can
be experimentally distinguished by examining the behavior of $|M_a|$ under field $H_c$.  

In a magnetic field, the phase transition from FE II to FE IV is first order, while the transition from FE IV to FE III is second order because $\theta_1$ and $\theta_2$ continuously change near the phase boundary. 
If as proposed, the electric polarization $\boldsymbol{P}$ in FE III is induced by $p$-$d$ orbital hybridization, then $\boldsymbol{P}\propto\sum_{i} (\boldsymbol{S}\cdot \boldsymbol{r}_i)^2 \boldsymbol{r}_i$ would lie along the $c$-axis ($\boldsymbol{r}_i$ represents the vector from the Fe atom to its nearest Cl atom or D$_2$O). If the $d$-$p$ orbital hybridization mechanism also holds in FE IV, then $\boldsymbol{P}$ would rotate away from the $c$-axis to the $a$-axis with a small angle ($<0.01 \pi$), causing a small $a$ component to coexist with a large $c$ component of the polarization.  Interestingly, this coexistence is observed in  $(\mathrm{NH}_4)_2\mathrm{FeCl}_5\cdot\mathrm{H}_2\mathrm{O}$ near $H_c=4$ T~\cite{Ackermann_2013}.

{\it Discussion and Conclusions.} --- We have studied the spin model proposed for $(\mathrm{ND}_4)_2\mathrm{FeCl}_5\cdot\mathrm{D}_2\mathrm{O}$ and examined the phase transitions under a magnetic field. We find that the second-order onsite spin anisotropy alone cannot describe the magnetic behavior of $(\mathrm{ND}_4)_2\mathrm{FeCl}_5\cdot\mathrm{D}_2\mathrm{O}$, including the weights of the third and fifth harmonic components of the cycloidal state and the appearance of the FE II state under a magnetic field along $c$-axis. 
With the fourth-order onsite spin anisotropy, the weights of the third and fifth harmonic components are enhanced and all the observed magnetic states of $(\mathrm{ND}_4)_2\mathrm{FeCl}_5\cdot\mathrm{D}_2\mathrm{O}$ are obtained in our simulations. Moreover, we predict a new FE IV state in the magnetic phase diagram. This state can be identified by measuring the uniform magnetization perpendicular to the magnetic field. Our results imply that the high-order onsite spin anisotropy is essential to explain the magnetic property of $(\mathrm{ND}_4)_2\mathrm{FeCl}_5\cdot\mathrm{D}_2\mathrm{O}$.

Our results qualitatively describe phase transitions of $(\mathrm{ND}_4)_2\mathrm{FeCl}_5\cdot\mathrm{D}_2\mathrm{O}$ under a magnetic field. To quantitatively describe the magnetic behavior, it would be necessary to carefully consider all parameters in the spin model, including the change of the exchange interactions under a magnetic field, to fine tune $K_2$ and $K_4$, and to include the spin-lattice interaction proposed in Ref.~\cite{AlbertoPRB2017}.

It is well known that the chemical substitution can change the spin anisotropy in solids and induce different magnetic ground states~\cite{WangPRB2018}. It would be interesting to study the effect of doping in $(\mathrm{ND}_4)_2\mathrm{FeCl}_5\cdot\mathrm{D}_2\mathrm{O}$, including the magnetic phase transitions and dynamical spin excitations. Our theoretical work on higher-order anisotropic interactions provides a guideline to understand the effect of doping. 

We would like to acknowledge useful conversations with Xiaojian Bai, Minseong Lee, Jan Musfeldt, and Wei Tian.
This work was supported by the U.S. Department of Energy, Office of Basic Energy Sciences, Materials Sciences and Engineering Division. 
This research used resources of the Compute and Data Environment for Science (CADES) at the Oak Ridge National Laboratory, which is supported by the Office of Science of the U.S. Department of Energy under Contract No. DE-AC05-00OR22725.

\bibliography{main}
\end{document}